\documentclass[twocolumn,preprintnumbers,floatfix,superscriptaddress,nofootinbib]{revtex4}
\pdfoutput=1 
\usepackage{CJKutf8} 
\usepackage{mathtools,slashed,mathrsfs,amsfonts}
\usepackage[caption=false]{subfig}
\usepackage{dcolumn}
\usepackage{multirow}
\usepackage{tabularx}
\usepackage{booktabs}
\usepackage{bm}
\usepackage{breqn}
\usepackage{comment}
\usepackage{setspace}
\usepackage[dvipsnames]{xcolor}
\usepackage[normalem]{ulem} 
\usepackage{enumerate}
\usepackage{siunitx}
\usepackage{multirow}
\usepackage{hyperref}

\newcommand{\MeV}{{\, {\rm MeV}}}
\newcommand{\GeV}{{\, {\rm GeV}}}

\definecolor{mypurple}{RGB}{164,64,214}

\newcommand\nn{\nonumber}
\newcommand\eea{\end{eqnarray}}
\newcommand\bea{\begin{eqnarray}}
\newcommand\ees{\end{split}}
\newcommand\bes{\begin{split}}

\newcommand{\Le}{\mathcal{L}_{\text{Entropy}}}
\newcommand{\Lt}{\mathcal{L}_{\text{Toy}}}
\newcommand{\Ll}{\mathcal{L}_{0}}

\def\l{\left(}
\def\r{\right)}

\begin{document}


\title{A Dynamical Explanation of the Dark Matter-Baryon Coincidence
}


\author{Dawid Brzeminski}
\email{dbrzemin@umd.edu}

\author{Anson Hook}
\email{hook@umd.edu}
\affiliation{Maryland Center for Fundamental Physics$,$ Department of Physics$,$ \\
University of Maryland$,$ College Park$,$ MD  20742$,$ U.S.A.}


\vspace*{1cm}

\begin{abstract} 

The near equality of the dark matter and baryon energy densities is a remarkable coincidence, especially when one realizes that the baryon mass is exponentially sensitive to UV parameters in the form of dimensional transmutation.  We explore a new dynamical mechanism, where in the presence of an arbitrary number density of baryons and dark matter, a scalar adjusts the masses of dark matter and baryons until the two energy densities are comparable.  In this manner, the coincidence is explained regardless of the microscopic identity of dark matter and how it was produced.  This new scalar causes a variety of experimental effects such as a new force and a (dark) matter density-dependent proton mass.

\end{abstract}

\maketitle

\section{Introduction}

The large hierarchy between the proton mass and the Planck scale was a cause of great consternation  back before QCD was first discovered. Eventually, a beautiful and elegant solution to this problem was discovered in the form of dimensional transmutation.  The QCD confinement scale, which determines the proton mass, is {\it exponentially} sensitive to UV parameters.  An $\mathcal{O}(1)$ number that happened to be $\sim 0.1$ is exponentiated giving the 18 orders of magnitude difference between the proton mass and the Planck scale.

The measured dark matter energy density ($\Omega_c = 0.26$) and baryonic energy density ($\Omega_b = 0.05$) are within a factor of $\sim 5$ of each other~\cite{Planck:2018vyg}.  This coincidence is extremely surprising given the extreme sensitivity of the proton mass (and hence $\Omega_b$) to $\mathcal{O}(1)$ numbers as well as the strong sensitivity of production mechanisms to various parameters.  Because of this, it is extremely surprising that $\Omega_b$ is within a factor of 5 of $\Omega_c$.  In this article, we seek to solve this coincidence problem.

Historically, the coincidence $\Omega_c \approx 5 \Omega_b$ has had only a single class of solution, whose general approach is as follows~\footnote{Anthropics can also be used to argue that under certain conditions that $\Omega_c$ is within a few orders of magnitude of  $\Omega_b$~\cite{Linde:1987bx,Wilczek:2004cr,Hellerman:2005yi}.}.  Firstly, a cosmological history is chosen such that the number densities of dark matter and baryons are approximately equal, $n_B \approx n_{DM}$.  Typically this is done via a $\mathbb{Z}_2$ discrete symmetry~\cite{Hodges:1993yb,Berezhiani:1995am} or by a shared asymmetry between dark matter and baryons~\cite{Nussinov:1985xr,Gelmini:1986zz,Barr:1990ca,Barr:1991qn,Kaplan:1991ah,Kaplan:2009ag}.  Secondly, the masses of the proton and dark matter are set to be approximately equal, $m_p \approx m_{DM}$.  This second step is typically ignored, but it can be accomplished by either a broken discrete symmetry~\cite{Foot:2003jt,An:2009vq,Farina:2015uea,GarciaGarcia:2015pnn,Lonsdale:2018xwd,Ibe:2019ena}, unification~\cite{Murgui:2021eqf}, coupled CFTs~\cite{Bai:2013xga,Newstead:2014jva,Ritter:2022opo} or sometimes simply by fiat.

In this article, we take a new approach to this coincidence problem.  We posit that the proton mass and the dark matter mass both depend on the expectation value of a scalar and when this scalar reaches its minimum energy configuration, it sets the total energy density of baryons and dark matter approximately equal.  While the details of the model we consider are given in Sec.~\ref{Sec: Toy} and Sec.~\ref{Sec: model}, it is simple to see how the mechanism works.  Due to dimensional transmutation, the proton and dark matter mass depend on a scalar $\phi$ as
\bea
m_p(\phi) = m_p(0) e^{c_B \phi/f} \quad m_{DM}(\phi) = m_{DM}(0) e^{-c_D \phi/f} . \nn
\eea
In the non-relativistic limit, the finite density potential for the scalar is
\bea
V(\phi) = m_p(\phi) n_B + m_{DM}(\phi) n_{DM}  \nn
\eea
so that at the minimum
\bea
\frac{d V(\phi)}{d\phi} = 0 \qquad \Rightarrow \qquad \frac{\rho_{DM}}{\rho_B} = \frac{c_B}{c_D} \sim \mathcal{O}(1) . \nn
\eea
As long as the potential for $\phi$ is dominated by the energy density of baryons $\rho_B$ and dark matter $\rho_{DM}$, it will relax the system to a state where the energy densities of the two are comparable, regardless of initial conditions.

Our new approach has many observational consequences.  Our mechanism does not just adjust $\rho_{DM} \sim \rho_B$ cosmologically but also inside of stars and planets, thereby giving the proton mass a dark matter density and normal matter density dependence.  This effect is so severe as to exclude the simplest implementation of the model forcing a small modification described in Sec.~\ref{Sec: model}.  Aside from a density-dependent proton mass, $\phi$ mediates an attractive fifth force between baryons, a repulsive new force between baryons and dark matter, and an attractive new force between dark matter.  

In Sec.~\ref{Sec: Toy} we present a toy model and its cosmology that highlights our mechanism.
We write down a Standard Model version of the toy model and elucidate its cosmology in Sec.~\ref{Sec: model}.  We discuss the constraints on our model in Sec.~\ref{Sec: bounds}.  Finally, we conclude in Sec.~\ref{Sec: conclusion}.

\section{Toy Model and Cosmology} \label{Sec: Toy}

We first discuss a toy model to highlight our relaxation mechanism and its associated cosmology.  The star of the show is a scalar $\phi$ that controls the mass of a baryon (B) and a dark matter (D) via dimensional transmutation.  We couple $\phi$ to our two sectors as
\bea \label{eq: toy}
\Lt = \frac{\phi}{f} \l \frac{\beta_B c_B}{32 \pi^2}  G_B^2 - \frac{\beta_D c_D}{32 \pi^2} G_D^2 \r.
\eea
$G_{B,D}$ are the field strength of confining sectors determining the mass of their respective particles and $\beta_{B,D}$ their beta functions.  
The linear coupling of $\phi$ to $G_{B,D}$ shifts their gauge couplings, and due to dimensional transmutation, this translates to a change in their corresponding confinement scales
\bea
m_B(\phi) &=& \Lambda_B(\phi) = \Lambda_B(0) e^{c_B \phi /f} \nn \\
m_D(\phi) &=& \Lambda_D (\phi) = \Lambda_D (0) e^{-c_D \phi /f}, \nn
\eea
where for simplicity, we have taken masses of the particles of interest equal to their confinement scale.  One could equally well consider other dark matter particles, e.g. axions, where their mass is proportional to the square of the dark confinement scale, $m_a \sim \Lambda_D (\phi)^2/f_a$.

\begin{figure}[t]
    \centering
    \includegraphics[width=0.49\textwidth]{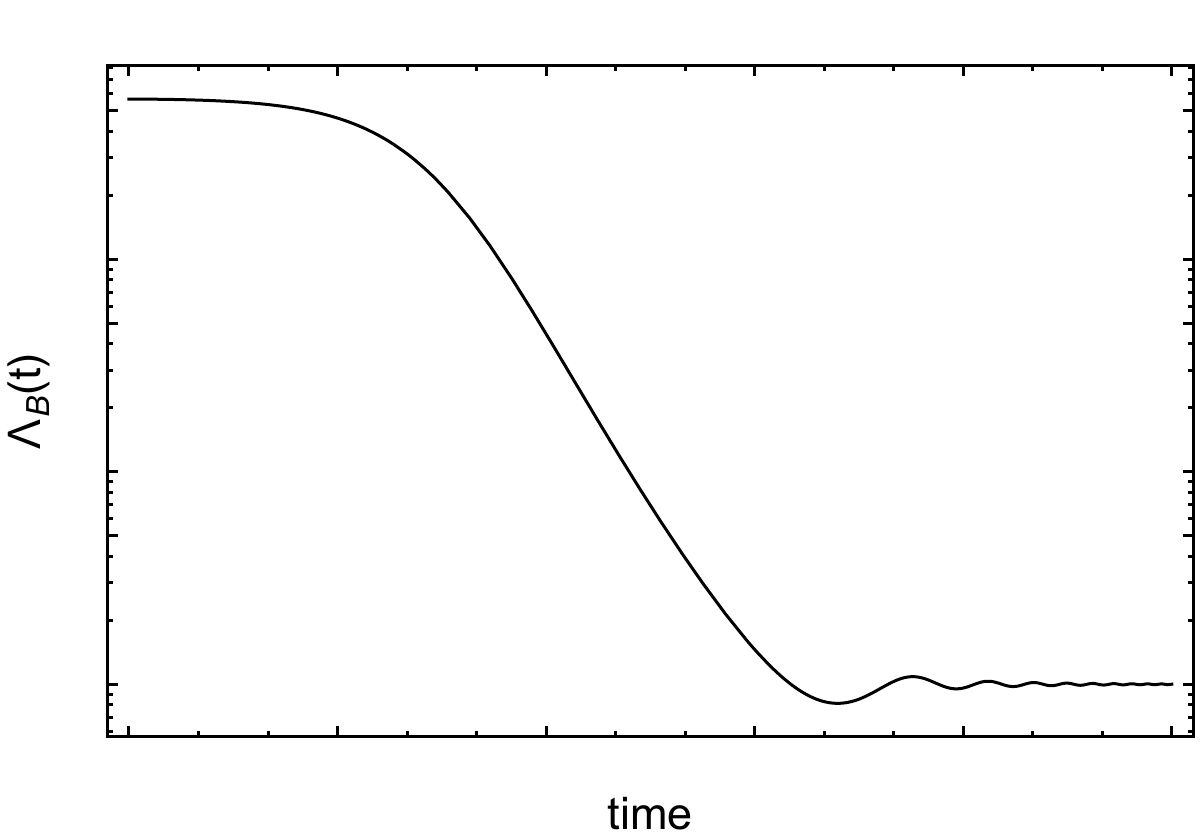}
    \caption{An example log-log plot of how the confinement scale evolves as a function of time in our toy model.  Initially, while $\phi$ is extremely overdamped, the confinement scale $\Lambda_B$ does not change.
After a while, a pseudo slow-roll with $m_\phi \sim H$ occurs, and the confinement scale relaxes with time.  Eventually, it reaches its minimum where $\rho_B \sim \rho_D$, and it starts to oscillate around the minimum with a rapidly decaying amplitude.
    }
    \label{fig: toy evoluation}
\end{figure}

We will be interested in the evolution of $\phi$ and the confinement scales as a function of time.   An example of how the confinement scale changes as a function of time is given in Fig.~\ref{fig: toy evoluation}.
The equation of motion we will be considering is
\bea \label{Eq: toy EOM}
\ddot \phi + 3 H \dot \phi = -\frac{c_B}{f}  m_B^0 e^{c_B \phi / f} n_B + \frac{c_D}{f}  m_D^0  e^{- c_D \phi/f} n_D.
\eea
Without the loss of generality, we shifted $\phi$ so that the minimum of the potential is located at $\phi = 0$.
We assume that baryons and dark matter have already been produced and are a non-relativistic subdominant energy density to whatever drives the expansion of the universe.
For simplicity, we will assume that the starting value of $\Lambda_B$ is larger than its final value so that $\Lambda_B$ ($\Lambda_D$) is relaxed to a smaller (larger) number.  In what follows, we specialize to a radiation-dominated universe only occasionally leaving comments on what changes as one varies the equation of state.

As is standard, the evolution of $\phi$ can be characterized by an underdamped and overdamped state.  Due to our assumption that baryons start off heavier than their final value, the mass of $\phi$ is dominated by the baryon's contribution giving
\bea
m_{\phi}^2 = V'' \approx \frac{c_B^2}{f^2}  m_B^0 e^{c_B \phi / f} n_B.
\eea
If initially $H > m_\phi$, then $\phi$ is frozen in place until $H \sim m_\phi$ where it enters a pseudo slow-roll regime.  Unlike the usual slow-roll scenario, $\ddot \phi$ and $\dot \phi$ terms in the equation of motion are equally important.  The mass of $\phi$ will change as a function of time, ensuring that $H \sim m_\phi$ for a prolonged period of time.  

The second scenario is the underdamped fast roll regime where $H > m_\phi$.  In this scenario, $\phi$ oscillates quickly in time.  Depending on the amplitude of the oscillation, the exponential form of the potential may play a critical role in changing how various energy densities dilute as a function of time.  Our analysis will be done in the limit where, despite the large oscillation of the QCD and dark confinement scales, dark matter and baryons both remain non-relativistic.

\paragraph{Pseudo Slow Roll} : 
If $m_\phi \ll H$, $\phi$ is essentially frozen in place.  Once $m_\phi \sim H$, $\phi$ to a good approximation follows the trajectory $a(t) \exp \l c_B \frac{\phi}{f} \r = \text{const}$. We can find the value of the constant by changing variables in the EOM to $x = \frac{a(t)}{a(t_i)} \exp \l c_B \frac{\phi}{f} \r$
\bea
\frac{\ddot{x}(t)}{x(t)} - \l \frac{\dot{x}(t)}{x(t)} \r^2 + 3 H(t) \frac{\dot{x}(t)}{x(t)}  - H(t)^2  \nn \\
+ \frac{c_B^2 m_B^0 n_B }{f^2} \l\frac{a(t)}{a(t_i)} \r^{-1} x(t)=0 ,
\eea
where we neglected the contribution from the dark sector.
Since we are looking for a static solution, we can neglect terms that involve derivatives, leading to a prediction
\bea
x(t) = \frac{f^2 H(t_i)^2}{c_B^2 m_B^0 n_B(t_i)} = \text{const.} \quad \Rightarrow \quad m_\phi(t) = H(t) .
\eea
From this, we see that $\Lambda_B(t) \sim a(t)^{-1}$, so that the confinement scale decreases like temperature~\footnote{For a matter-dominated universe relaxation is a very slow occurrence as to leading order $\Lambda_B(t) \sim a(t)^0$, while for a kination dominated universe $\Lambda_B(t) \sim a(t)^{-5}$.}.
Eventually, $\phi$ nears its minimum, and the approximation of neglecting dark matter's contribution to the $\phi$ equation of motion fails.
After crossing the minimum of the potential at $\phi = 0$, $\phi$ oscillates as an underdamped harmonic oscillator, as described in the next subsection. 

We next estimate the kinetic energy of $\phi$ during pseudo slow-roll.  We use the relation
\bea \label{eq: dphi in x}
\dot{\phi} = \frac{f}{c_B} \l \frac{\dot{x}}{x} - H(t) \r \approx - \frac{f H(t)}{c_B} ,
\eea
neglecting the small $\frac{\dot{x}}{x}$ term to estimate the kinetic energy of $\phi$ as
\bea
\frac{1}{2} \dot \phi^2 = \frac{f^2 H(t)^2}{2 c_B^2} = \frac{m_B(t) n_B(t)}{2}
\eea
from which we see that during our pseudo slow-roll the kinetic energy in $\phi$ is comparable but slightly subdominant to the energy density in baryons.

\paragraph{Underdamped regime} :
When $m_\phi \gtrsim H$, $\phi$ oscillates quickly around its minimum.  The energy in $\phi$ redshifts differently  depending on if the amplitude of the oscillation obeys $\phi \lesssim f/c_B$ or $\phi \gtrsim f/c_B$.

For small amplitudes, as is the case when one transitions into this regime from pseudo slow-roll, $\phi$ behaves as a harmonic oscillator with a time-dependent mass.  As is standard, using the WKB approximation we find $\rho_{\phi} \propto a^{-9/2}$ coming from a combination of $m_{\phi} \propto a^{-3/2}$ and $n_\phi \sim a^{-3}$.

For large amplitudes, we find numerically that $\rho_{\phi} + \rho_{B} \propto 1/a^{5}$ while the amplitude of the oscillating confinement scale decays as $\Lambda_B \sim 1/a^{2}$.  As this behavior results in wildly oscillating confinement scales, potentially invalidating the non-relativistic approximation, we will not consider this limit further.

\section{Standard Model example and cosmology} \label{Sec: model}

There are two main differences when generalizing the previous example to the Standard Model.  
The first difference is that it is not possible to only adjust the QCD scale as loop effects will result in $\phi$ adjusting other things such as electric charge.  At the loop level, the SM plasma energy density depends on the fine structure constant, and this gives a large temperature-dependent potential for $\phi$ that is potentially larger than the contribution coming from baryons, see the Appendix for a more quantitative description of this problem.
We will solve this problem with an entropy dump. 
The second difference is that the model, as previously written, is excluded by measurements today as it gives the proton an unacceptably large density dependence.  
This problem is solved by a combination of a bare potential with the aforementioned entropy dump.

The Lagrangian that we consider is
\bea \label{eq: effective lagrangian}
\mathcal{L} = \Lt + \Ll + \Le.
\eea
$\Lt$ is given by Eq.~\ref{eq: toy} with the baryonic confinement scale replaced by the QCD scale.  $\Ll$ is a potential for $\phi$
\bea \label{Eq: late}
\Ll = \Lambda_0^4 \cos \l \frac{\phi}{F} + \theta \r .
\eea
Meanwhile, our entropy dump can take any form, but for concreteness, we take
\bea
\label{Eq: entropy}
\Le = \kappa \Phi_E H H^\dagger ,
\eea
where a heavy (heavier than a TeV) mass scalar $\Phi_E$ reheats the SM via decays into the Higgs boson.

\begin{figure}[t]
    \centering
    \includegraphics[width=0.49\textwidth]{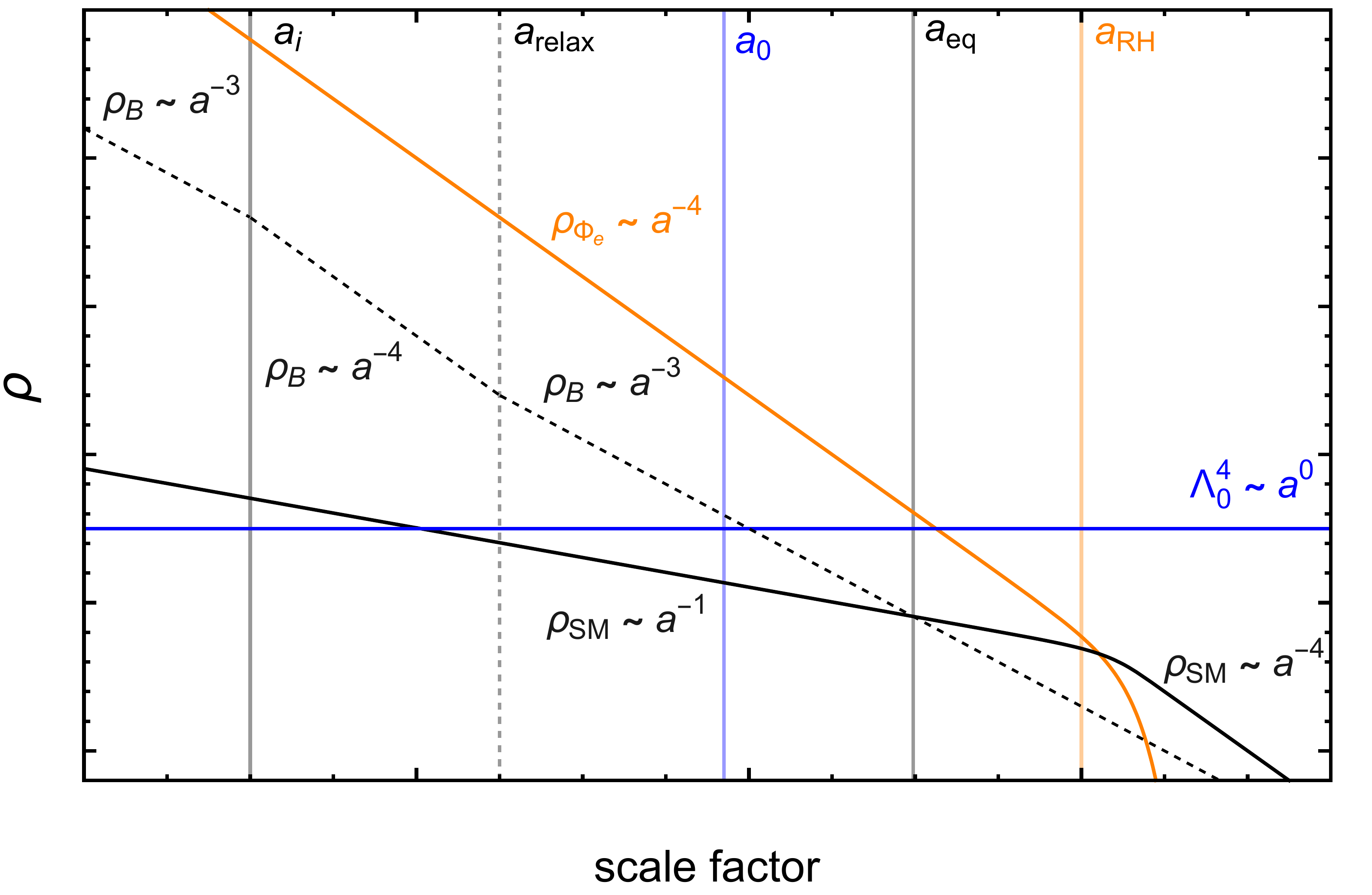}
    \caption{An example of the cosmological history we are interested in is shown on a log-log scale.  In orange, we have the energy density of the relativistic particle $\Phi_E$, which decays at a scale factor $a_{RH}$, reheating the Standard Model and providing an entropy dump.  In blue, we have the bare potential for $\phi$ which becomes important when $\Lambda_0^4 \sim \frac{1}{2} \dot \phi^2 \lesssim \rho_B$.  In the dashed black line, we have the energy density in baryons.  Initially, it decays away as $a^{-3}$.  At a scale factor, $a_i$, $m_\phi \sim H$ and it decays as $a^{-4}$ as the QCD scale is being adjusted to its final value.  At a scale factor $a_{\text{relax}}$, the QCD scale reaches its final value while at the scale factor $a_0$, $\phi$ falls into its nearest density independent minimum.
    Finally, in a solid black line, there is the thermal energy density of the Standard Model.  Initially, the energy density falls as $a^{-1}$ as it is being constantly replenished by early decays of $\Phi_E$.  At a scale factor $a_{\text{eq}}$, its energy density overtakes that of the baryons.  Finally, after reheating, it falls like radiation.}
    \label{fig: evoluation}
\end{figure}

A pictorial representation of our cosmological history is shown in Fig.~\ref{fig: evoluation}.  There are four energy densities that are important.  Firstly, there is the relativistic species $\Phi_E$, which dominates the energy density at early times and whose decays reheat the Standard Model and provide an entropy dump.  Secondly, there is $\Lambda_0^4$, which is the bare potential for $\phi$.  Thirdly, there is the energy density in baryons $\rho_B$, which dilutes like radiation while the QCD scale is being scanned and later decays away like matter.  Finally, there is the thermal energy density in the SM that is not in the rest mass of the baryons, $\rho_{SM}$.

Let us first give a word-level explanation of our cosmology.  Initially, $m_\phi \ll H$ and no relaxation is taking place.  At a scale factor $a_i$, the QCD scale begins to be adjusted in a radiation-dominated universe so that $\Lambda_{QCD} \sim 1/a(t)$.  Eventually, at a scale factor $a_{\text{relax}}$, the QCD scale reaches its minimum where $\rho_D \approx 5 \rho_B$.  At a scale factor $a_0 \gtrsim a_{\text{relax}}$, the $\Lambda_0^4$ potential becomes important, preventing any further relaxation and fixing $\rho_D \sim 5 \rho_B$ in stone.  A little bit later, at $a_{eq} \gtrsim a_0$, the potential from the thermal bath of the SM becomes equal to that of the baryons.  Finally, at a scale factor $a_{RH}$, the entropy dump concludes, and the SM is fully reheated and becomes the dominant energy density in the universe.

In what follows, we give a more detailed description of the cosmology.  For simplicity, we will discuss our cosmological history in the context of a single data point while being slightly cavalier about $\mathcal{O}(1)$ numbers.  The scale factors will be related to each other as $a_i = 5 \times 10^{-8} a_{RH}$, $a_{\text{relax}} = 5 \times 10^{-6} a_{RH}$, $a_0 = 2 \times 10^{-4}$ and $a_{eq} = 2 \times 10^{-4} a_{RH}$.  
We will take $T_{RH} = 10$ MeV.  The initial value of the proton mass is 100 GeV 
and relaxes to 1 GeV~\footnote{If the initial proton mass was larger than 100 GeV but with all other parameters unchanged, then everything would proceed exactly as in our data point, just with $a_i$ slightly smaller.}.  
In order for relaxation to start at $a_i$, $m_\phi = H$ at that time gives $f/c_B \sim 10^{12}$ GeV.  With an eye on constraints and since $F$ has to be small enough to scan the QCD scale, we take $F \sim 10^{7}$ GeV.
Finally, we take $\Lambda_0 \sim 100$ MeV.  A summary of some of the energy densities and their values at different times is shown in Tab.~\ref{Table}. 

\paragraph{$\rho_{\Phi_E}$} : We take as our initial conditions a universe dominated by a relativistic species $\Phi_E$.  This can occur in a number of ways, e.g., if the inflaton decayed into $\Phi_E$.  It is important that $\Phi_E$ is relativistic so that its energy density dilutes away as radiation since the QCD scale is relaxed as $\Lambda_{QCD} \sim 1/a$ ($\Lambda_{QCD} \sim$ const.) in a radiation (matter) dominated universe.

We take the mass of $\Phi_E$ to be $M_{\Phi_E} = 10$ TeV.  Given a boost $\gamma$, its lifetime is 
\bea
\Gamma = \frac{\kappa^2}{4 \pi M_{\Phi_E} \gamma} .
\eea
The universe is reheated when $\Gamma(t) = H(t)$ at a scale factor $a_{RH}$.  At this time, the energy density in $\Phi_E$ decays away, and the SM becomes the dominant energy density in the universe.  For our choice of parameters, the reheating temperature of the universe is $T_{RH} = 10$ MeV with $\kappa \sim 10^{-7}$ GeV and $\gamma = 10^3$ at decay.

\paragraph{$\Lambda_0^4$} : The bare potential for $\phi$ is always present.  It becomes important when the kinetic energy of $\phi$ is no longer able to take it over the $\Lambda_0^4$ sized barrier.  This can occur during relaxation when $\Lambda_0^4 \sim \frac{1}{2} \dot \phi^2 \sim \rho_B$, or it can occur afterwards as the kinetic energy of $\phi$ rapidly redshifts away.  As long as $2 \pi F \lesssim f/c_B$, then there will be a minimum close by where $\rho_D \sim 5 \rho_B$.

\paragraph{$\rho_{B}$} : The energy density in baryons behaves in a manner described in the previous toy model.  Initially, it is diluting away as matter when $m_\phi \ll H$.  At $a_i$ the QCD scale starts to relax as $1/a(t)$ so that $\rho_B \sim 1/a(t)^4$.  Relaxation finishes at a scale factor $a_{\text{relax}}$, and afterwards the baryons dilute like normal non-relativistic matter $\rho_B \sim 1/a(t)^3$.

\paragraph{$\rho_{D}$} : 
Dark matter has more freedom than baryons as it is not important to the dynamics of $\phi$ until $a_{\text{relax}}$.  The simplest cosmology involves dark matter already being present before $a_i$.  Initially, dark matter dilutes away as $a^{-3}$.  At $a_i$ the dark confinement scale rises as $a^{c_D/c_B} \sim a^{0.2}$ so that $\rho_D \sim 1/a(t)^{2.8}$.  Again, at $a_{\text{relax}}$, dark matter returns to diluting away as cold dark matter.

\begin{table}[t] 
\centering
 \begin{tabular}{||c | c c c||} 
 \hline
  & initial & relax & reheat \\ [0.5ex] 
 \hline\hline
 a & $5 \times 10^{-8} \, a_{RH}$ & $5 \times 10^{-6} \, a_{RH}$ & $a_{RH}$ \\ 
 $\rho_{\Phi_E}$ & $5\times 10^9 \, \text{TeV}^4$ & $ 50 \, \text{TeV}^4$ &  $\l 10 \, \text{MeV} \r^4$ \\
 $\rho_{SM}$ & $3 \times 10^{-1} \, \text{GeV}^4$ & $3\times 10^{-3} \, \text{GeV}^4$ &  $\l 10 \, \text{MeV} \r^4$ \\
 $m_p$ & $100 \, \text{GeV}$ &  $1 \, \text{GeV}$ & $ 1\,\text{GeV}$ \\
 $n_B$ & $3 \times 10^6 \, \text{GeV}^3$  & $ 3 \, \text{GeV}^3$ & $4 \times  10^{-7}  \, \text{MeV}^3$ \\ [1ex] 
 \hline
 \end{tabular}
 \caption{Some of the relevant quantities for our example data point at the three interesting times:  when the QCD scale starts to relax, when it stops relaxing, and when the Standard Model is reheated.  }\label{Table}
\end{table}

\paragraph{$\rho_{SM}$} :  The last piece of the puzzle is the temperature of the SM.  As described in the Appendix, a non-zero temperature of the SM gives a temperature-dependent potential that scales as $10^{-6} \, T^4$ (or even larger if particles such as pions are present) that prevents relaxation if it is larger than $\rho_B$.  As such, we require that all relaxation occurs before $\rho_{SM} \sim \rho_B$ giving $a_{eq} \gtrsim a_{\text{relax}}$.

We take as initial conditions that $\rho_{SM} \ll \rho_B$.  This can be obtained through extremely efficient baryogenesis models such as Affleck-Dine baryogenesis where $Y_B \sim 10^3$~\cite{Affleck:1984fy} or simply by allowing for some red shifting so that matter eventually overtakes radiation.

The thermal energy density in the SM cannot be set arbitrarily small as early decays of $\Phi_E$ will give the SM a minimal temperature.  This temperature is easily solved for using energy conservation, but a simple parametric estimate can be obtained using the energy deposited into the SM during a single Hubble time
\bea
\rho_{SM}(t) \sim \frac{\Gamma(t)}{H(t)} \rho_{\Phi_E} ,
\eea
giving an energy density that falls in time as $1/a(t)$.

\section{Bounds} \label{Sec: bounds}

\subsection{Neutron Star}

While the potential in Eq.~\ref{Eq: toy EOM} works in the early universe, it will fail today as any macroscopic body is many orders of magnitude denser than the average cosmological density.  As a result, we would end up with much lighter protons inside macroscopic bodies than in the vacuum.  The most extreme example that we need to clear is the neutron star, $\rho_{NS} \sim (100 \MeV)^4$. Taking into account the bare potential for $\phi$, the potential inside of a neutron star is approximately
\bea
V = \rho_{NS} \exp \l c_B \l  \frac{\phi}{f}  \r\r + \Lambda_0^4 \cos \l \frac{\phi}{F} + \theta \r
\eea
Requiring that the neutron star does not displace the minimum by more than $F$  gives
\bea \label{eq: astro bound}
\frac{\Lambda_0^4}{F} &\gtrsim& \frac{c_B}{f} \rho_{NS} .
\eea
This bound is satisfied for the parameters we chose.

\subsection{Fifth force and Stellar Constraints}

With any light Yukawa coupled scalar, the dominant constraints come from stars and fifth force measurements.
$\phi$ mediates a long-range force between nuclei
\bea
V &\supset& \exp \l \frac{c_B \phi}{f} \r m_{B} \bar{\psi} \psi \rightarrow  \frac{c_B \phi}{f} m_{B} \bar{\psi} \psi .
\eea

The example data point we chose barely squeaks by the current stellar constraints~\cite{Hardy:2016kme}.
Evading fifth force measurements requires this force to have to have a short enough range.  Our $\phi$ mass is
\bea \label{eq: mass bound}
m_{\phi} &=& \frac{\Lambda_0^2}{F} = 1 \, \text{eV} \l \frac{F}{10^{7} \GeV} \r^{-1},
\eea
and was chosen to be right on the edge of fifth force constraints~\cite{Hoskins:1985tn,Smith:1999cr,Kapner:2006si,Decca:2007jq,Schlamminger:2007ht,Geraci:2008hb,Sushkov:2011md,Lee:2020zjt}.

\subsection{$\phi$ decays}

Amusingly, $\phi$ can very easily be a significant fraction of dark matter as its kinetic energy during pseudo slow-roll is comparable to the baryonic energy density.
As discussed in the Appendix, $\phi$ is necessarily coupled to photons and thus can decay into them.  While in our particular data point, the $\phi$ lifetime is long enough that it evades current constraints, in some regions of parameter space this is a concern~\cite{Bernal:2020lkd,Todarello:2023hdk}.

\section{Conclusions} \label{Sec: conclusion}

In this article, we presented a new approach towards explaining $\Omega_c \approx 5 \Omega_b$.  A scalar adjusts the baryon and dark matter masses such that $\Omega_c \approx 5 \Omega_b$ regardless of the production mechanism for baryons and dark matter and independent of the identity of dark matter.  This approach thus allows one to explain why promising dark matter candidates, such as the QCD axion, have energy density so close to that of the baryons despite their production mechanisms being completely independent.

Our adjustment mechanism is extremely testable.  The scalar $\phi$ mediates a new force, potentially visible in fifth force experiments and astrophysical environments.  It also gives the proton mass such a strong environmental dependence that modifications needed to be made to evade this constraint.

What we presented is the essence of a new mechanism.  It still remains to explore all regions of parameter space.  Additionally, this mechanism is extremely versatile and it would be interesting to see how it meshes with various favored dark matter candidates.  Finally, while the EFT is extremely minimal, the UV completion is more complicated.  It would be exciting if a more compelling UV completion were to be put forth.

\section*{Acknowledgments}

We thank Asimina Arvanitaki and Junwu Huang for useful discussions.  DB and AH are supported in part by the NSF grant PHY-2210361 and the Maryland Center for Fundamental Physics.
This work was completed in part at the Perimeter Institute and the Aspen Center of Physics.  Research at Perimeter Institute is supported in part by the Government of Canada through the Department of Innovation, Science and Economic Development Canada and by the Province of Ontario through the Ministry of Colleges and Universities. The Aspen Center for Physics is supported by National Science Foundation grant PHY-2210452.

\bibliography{DM_B}{}

\bibliographystyle{JHEP}

\clearpage
\newpage
\maketitle
\onecolumngrid
\begin{center}
\textbf{\large Dynamical Equilibration of Dark Matter and Baryon Energy Densities} \\ 
\vspace{0.05in}
{ \it \large Supplementary Material}\\ 
\vspace{0.05in}
{}
{Dawid Brzeminski, Anson Hook}

\end{center}
\setcounter{equation}{0}
\setcounter{figure}{0}
\setcounter{table}{0}
\setcounter{section}{0}
\renewcommand{\theequation}{S\arabic{equation}}
\renewcommand{\thefigure}{S\arabic{figure}}
\renewcommand{\thetable}{S\arabic{table}}
\newcommand\ptwiddle[1]{\mathord{\mathop{#1}\limits^{\scriptscriptstyle(\sim)}}}

This Supplementary Material contains additional calculations supporting the results in the main text.

\section{UV completion} 
 \label{appendix: UV}
 
In this section, we discuss the UV completion of Eq.~\ref{eq: effective lagrangian}.  We will focus on Eq.~\ref{eq: toy} in order to explain the exponential coupling of $\phi$ to the QCD or dark scale.  
The other two terms in Eq.~\ref{eq: effective lagrangian} have standard UV completions and will not be discussed further.  
The last aspect of Eq.~\ref{eq: effective lagrangian} is the lack (smallness) of any other potential.  This problem has been solved before and we refer the reader to Refs.~\cite{Hook:2018jle,Brzeminski:2020uhm} for a detailed explanation.

We now discuss how to generate an exponential coupling of $\phi$ to the QCD (dark) scale.  We begin by introducing $N_\chi$ heavy vector-like Dirac fermions $\chi$ in the fundamental representation of the confining sector.  $\chi$ has both a bare mass and a Yukawa coupling to $\phi$
\bea \label{eq: Interaction Lagrangian}
\mathcal{L}_I = \l M + y \phi \r \bar{\chi} \chi .
\eea
As a result, the mass of this vector-like fermion depends on $\phi$
\bea
M_\chi = M \l 1+ \frac{ y \phi}{M} \r .
\eea

$\chi$ affects the IR value of the strong coupling as
\bea
\Delta \l \alpha^{-1} \r = - \frac{N_\chi}{3 \pi} \log \l \frac{M_\chi(\phi)}{\Lambda_{UV}} \r
\eea
which at 1-loop translates to a $\phi$-dependence of the confinement scale
\bea
\Lambda(\phi) = \Lambda(0) \l 1+  \frac{y \phi}{M} \r^{\frac{2 N_{\chi}}{3 \beta}} \approx \Lambda(0) \exp \l c_B \phi/f \r
\eea
where $c_B/f = \frac{2 y  N_{\chi}}{3 \beta M}$ and $\beta = 11 N_c/3 - 2 N_f/3$ is the beta function not including $\chi$.  We have taken the large $N_\chi$ and small $y$ limit to ensure the exponential form.  Numerically, we find that our results hold even without the exponential approximation.

By utilizing this UV completion for both the dark sector and QCD, an exponential coupling of $\phi$ to the masses of the proton and dark matter is obtained.  The only other minor point is that $\phi$ needs to be made periodic in order to suppress its bare potential to acceptable levels, ala Refs.~\cite{Hook:2018jle,Brzeminski:2020uhm}.  This can be accomplished by making $\phi$ the phase of a complex scalar $\Phi$ that undergoes spontaneous symmetry breaking.  The model in this section only needs to be modified by replacing $\phi \Rightarrow \Phi, \Phi^\dagger$.

\section{Thermal potential for $\phi$}

In this section, we discuss how despite $\phi$ only scanning the QCD gauge coupling, loop-level effects can lead to additional potentials for $\phi$ that are independent of the baryon number density.  In what follows, we will be somewhat qualitative as many of the results depend on small details such as the number of electrically charged particles present in the thermal bath.

\subsection{Pions}

Aside from baryons, pions also have a QCD scale dependent mass, $m_\pi \sim \sqrt{(m_u + m_d) \Lambda_{QCD}}$.  If the temperature of the SM is between $m_\pi \lesssim T \lesssim 4 \pi \Lambda_{QCD}$, then the pions give a temperature-dependent contribution to the potential of $\phi$ that scales like
\bea
V(\phi) \approx \frac{1}{8} m_\pi^2 T^2 + \cdots.
\eea
When $T \sim m_\pi$, this potential for $\phi$ can reach a strength as large as $T^4$ in magnitude.

\subsection{Fine structure constant}

The next contribution to the potential of $\phi$ that we consider is a two-step problem.  Firstly, due to loop effects, a scalar $\phi$ that scans the QCD scale will inevitably scan the fine structure constant as well.  Secondly, due to loops of charged particles, there is always a contribution to the potential energy of a system in the form of $e^2 T^4$ as long as $T$ is larger than the mass of a charged particle.

First, we discuss how loop effects cause $\phi$ to also scan the fine structure constant.  Let us first write down the two couplings we are interested in
\bea
\mathcal{L} = \frac{\phi}{f} \frac{\beta_B c_B}{32 \pi^2}  G^2 + c_\gamma \frac{\phi}{f}F^2 .
\eea
We are interested in what $c_\gamma$ is generated at low energies, assuming that it is 0 in the UV.  At two loops, going through a loop of quarks and gluons, one can already see that a coupling to photons is generated.  This is even more complicated by the fact that we need to RG evolve through the strongly coupled QCD phase transition so that we expect any calculation to be $\mathcal{O}(1)$ uncertain.

An estimate for the coupling to photon generated by integrating through the QCD scale can be obtained by realizing that to leading order in $1/f$, our coupling to QCD is the same as a dilaton.  Since the dilaton always couples as the beta function, we know that integrating through the QCD scale changes $c_\gamma$ by the difference between the UV and IR beta functions.  Thus we can estimate the size of $c_\gamma$  from transitioning from 3 flavor QCD to a theory of pions and kaons as 
\bea
\delta_{QCD} c_\gamma \approx \frac{c_B}{32 \pi^2} \l \beta_\gamma^{\pi,K} - \beta_\gamma^{\text{3 quark}} \r = \frac{c_B}{16 \pi^2} .
\eea
Afterwards, RG evolution of $\alpha$ coming from charged pions and kaons also changes
\bea
\Delta \frac{1}{\alpha} = - \frac{1}{6 \pi} \log \frac{m_\pi}{\Lambda_{QCD}}  - \frac{1}{6 \pi} \log \frac{m_\pi}{\Lambda_{QCD}} \sim \frac{1}{6 \pi} \log \Lambda_{QCD} \sim \frac{c_B \phi}{6 \pi f} \qquad \Rightarrow \qquad 
\delta_{RG} c_\gamma \approx \frac{c_B}{96 \pi^2} .
\eea
Combined we arrive at the conclusion that
\bea
c_\gamma \sim \mathcal{O}(1) \frac{c_B}{32 \pi^2} .
\eea

The next step is to estimate how the free energy depends on the fine structure constant, see e.g. Refs.~\cite{Laine:2016hma,Brzeminski:2022haa} for how to do such a computation.  While the final answer depends on what particles are in the thermal bath, the two-loop contribution coming from the electron is
\bea
f = \frac{5}{288} e^2(\phi) T^4 \sim  10^{-6}  \, T^4 \, \frac{c_B \phi}{f} ,
\eea
where we have done a Taylor series for small $\phi$.  While highly suppressed, if one compares this term in the potential to the one due to $\rho_B$, then at 10 MeV with a normal cosmology, one finds that this term is more important by a few orders of magnitude.  As a result of this issue, we were forced to consider an entropy dump.

\section{Parametric Resonance}

As $\phi$ scans the QCD scale, it transverses a large distance in field space and in the process passes over the bumpy $\Lambda_0^4$ sized cosine potential.  This process can cause parametric resonance whereby a non-zero $k$ mode of $\phi$ becomes unstable removing energy from the zero mode and eventually causing $\phi$ to fragment, see e.g. Ref.~\cite{Fonseca:2019ypl} for a detailed study in a somewhat similar situation.  Up to a logarithmic correction, the timescale associated with this fragmentation is
\bea
t_{\rm frag} \sim \frac{F \dot \phi^3}{\Lambda_0^8} \sim \frac{F \rho_B^{3/2}}{\Lambda_0^8} .
\eea
In order for fragmentation to never be important in stopping the movement of $\phi$, we require that at the scale factor $a_{\text{relax}}$ that $H \, t_{\rm frag} \gtrsim 1$ roughly giving the bound
\bea
\frac{F \rho_B^{3/2} \rho_{\Phi_E}^{1/2}}{\Lambda_0^8 M_{pl}} \gtrsim 1 \qquad {\rm when} \qquad a = a_{\text{relax}}.
\eea
This constraint can be avoided at the expense of some additional model building complexity.  If $\Lambda_0$ depends on temperature, in much the same way as the QCD axion potential depends on temperature, then the $\Lambda_0^4$ wiggles need not be present at early times and the bound discussed in this section would be evaded.

\end{document}